\documentstyle[aps,amssymb,psfig]{revtex}
\begin{document}

\author{
Simonetta Frittelli$^a$ \and
   Ezra T. Newman$^{b}$ \\
   $^{a}$Department of Physics, Duquesne University,
       Pittsburgh, PA 15282\\
   $^{b}$Department of Physics and Astronomy,
       University of Pittsburgh,
       Pittsburgh, PA 15260
}
\title{
\rightline{\small{\em To appear in Phys. Rev. D \/}}
Dynamics of Fermat potentials in non-perturbative gravitational
    lensing}
\date{April 12, 2002}
\maketitle

\begin{abstract}

We present a framework, based on the null-surface formulation of general
relativity, for discussing the dynamics of Fermat potentials for gravitational
lensing in a generic situation without approximations of any kind.
Additionally, we derive two lens equations: one for the case of thick compact
lenses and the other one for lensing by gravitational waves. These equations in
principle generalize the astrophysical scheme for lensing by removing the
thin-lens approximation while retaining the weak fields.

\end{abstract}

\section{Introduction}\label{sec:1}

In the astrophysical approach to lensing~\cite{EFS,pettersbook}, background
light sources are considered to lie on a plane at a distance $D_s$ from the
observer, and the deflector or lens is described by a mass distribution on a
plane at distance  $D_l$ from the observer. Lightrays from the source in the
source plane travel along straight null geodesic paths on a flat or
cosmological background until they reach the lens plane, at which point they
suffer a sharp bend in direction, subsequently traveling again on a straight
line towards the observer.  The direction at which a lightray reaches the
observer determines the angular location of the image on the observer's
celestial sphere. The celestial sphere can be identified with the lens plane
(which is sometimes referred to as the image plane). A fundamental element of
gravitational lensing is a lens equation of the form:

\begin{equation}\label{lenseq}
    \bbox{\eta} = \frac{D_s}{D_l} \bbox{\xi} - D_{ls} \bbox{\alpha}(\bbox{\xi})
\end{equation}

\noindent where $\bbox{\eta}$ is a position vector on the source plane representing
the source's location, $\bbox{\xi}$ is a position vector on the lens plane
representing the location where the lightray emitted by the source pierces the
lens plane, $D_{ls}$ is the distance between the lens plane and the source
plane and $\bbox{\alpha}$ is the bending angle that the lightray suffers at the lens
plane, which is determined by the distribution of mass on the lens plane and is
thus a function on the lens plane.  The bending angle $\bbox{\alpha}$ can be obtained
from a deflection potential $\psi(\bbox{\xi})$ via

\begin{equation}
    \bbox{\alpha} = \frac{\partial \psi}{\partial \bbox{\xi}}
\end{equation}

\noindent and the deflection potential $\psi$ in turn is obtained from the mass
distribution $\Sigma(\bbox{\xi})$ by means of a 2-dimensional Poisson equation

\begin{equation}\label{eq:0}
    \Delta_{\bbox{\xi}}\psi(\bbox{\xi}) = \Sigma(\bbox{\xi}),
\end{equation}

\noindent where $\Delta_{\bbox{\xi}}$ is the 2-dimensional Laplacian on
the lens plane.

A very elegant scheme that leads to the lens equation is based on Fermat's
principle, according to which light takes the path of least travel time between
any two fixed points.  In order to use this principle, a set of paths between
the fixed source at $\bbox{\eta}$ and the observer are considered, all of which
consist of two straight segments, but which are allowed to pierce the lens
plane at arbitrary points $\bbox{\xi}$.  The value of $\bbox{\xi}$ thus becomes a label for
the path.  The travel time along such paths is thus a function of $\bbox{\xi}$, which
must be minimized with respect to $\bbox{\xi}$.  The travel time along such paths is a
Fermat potential $\phi(\bbox{\xi},\bbox{\eta})$, in general depending on both the source point
as well as the point on the lens plane.  The Fermat potential has the form

\begin{equation}
    \phi(\bbox{\xi},\bbox{\eta})
   =
    \frac12\frac{D_lD_s}{D_{ls}}
    \left|\frac{\bbox{\eta}}{D_s} - \frac{\bbox{\xi}}{D_l}\right|^2
    -\psi(\bbox{\xi}),
\end{equation}

\noindent where the first term represents the geometric length of the broken
path with respect to the length of a straight path between the source and the
observer, and the second term represents the gravitational time delay suffered
by the lightray as a consequence of the presence of mass at the lens plane. The
minimization of the travel time with respect to the path requires

\begin{equation}
    \frac{\partial\phi(\bbox{\xi},\bbox{\eta}) }{\partial \bbox{\xi}}= 0.
\end{equation}

\noindent This equation provides the lens map $\bbox{\xi} \to \bbox{\eta}$ from the lens
plane to the source plane, i.e., Eq~(\ref{lenseq}).  The Fermat potential is
affected by the presence of a mass distribution $\Sigma(\bbox{\xi})$ on the lens plane
essentially because of Eq.~(\ref{eq:0}).  As a consequence of Eq.~(\ref{eq:0}),
one can see that the Fermat potential also satisfies a Poisson-type equation

\begin{equation}
\Delta_{\bbox{\xi}}\phi(\bbox{\xi},\bbox{\eta}) = -S(\bbox{\xi}),
\end{equation}

\noindent where  $S(\bbox{\xi})$ differs from $\Sigma(\bbox{\xi})$ by an additive constant.
This is the dynamical equation for the Fermat potential in the astrophysical
approach to lensing. It encodes the information of the curvature of the
spacetime due to the presence of mass on the lens plane, and it thus represents
the role of the Einstein equations in lensing.  One can see that the
Fermat-potential formulation of lensing is fully tailored to the approximations
in use in the astrophysical approach to lensing, namely: thin lenses, weak
fields and small angles.

We are interested in a generalization of the Fermat-potential approach to
lensing for a generic situation with no approximations. The first step to this
goal is to construct a Fermat potential for a generic situation, such that a
lens mapping from the observer's celestial sphere to the lightsource locations
results from a variational principle applied to it. We refer to this phase of
the project as the kinematics of the Fermat potential.  The second step is to
find a ``field equation'' for the Fermat potential, namely: an equation that
relates the Fermat potential to the sources of gravitational field. We regard
this as the dynamics of the Fermat potential.

In a previous work~\cite{fermatkling} we have described a setting
for an approach to the kinematics of Fermat potentials using both a
generalization of a Fermat potential and a generalized Fermat principle
consistent with a generic situation, which we briefly summarize in the
following.

The aim of the scheme is to construct a function on spacetime that, by
extremization, leads to an expression of the observer's past lightcone. The
observer's past lightcone can be parametrized in terms of the celestial sphere
(the null directions at the apex of the lightcone) and an affine length along
the null rays that rule the lightcone.  Such a parametrization leads directly
to a lens mapping from the celestial sphere out to lightsources, and a
time-of-arrival equation for the light signal reaching the observer. Thus, our
immediate aim is to provide a scheme by which the observer's past lightcone can
be constructed. Since lightcones are level surfaces of solutions to the eikonal
equation, such solutions are our starting point.

Consider a lensing spacetime with coordinates $x^a$ and metric $g_{ab}$.
Suppose we have a (sufficiently generic) 2-parameter family of solutions
$Z(x^a,\zeta,\bar\zeta)$ of the eikonal equation $g^{ab}Z,_aZ,_b=0$ (with $,_a
\equiv \partial/\partial x^a$), such that the surfaces of constant value of
$Z(x^a,\zeta,\bar\zeta)$ foliate the spacetime for every fixed value of
$(\zeta,\bar\zeta)$.  Define

\begin{equation}\label{eq:1}
    G(x_0^a,x^a,\zeta,\bar\zeta)
    \equiv
    Z(x^a,\zeta,\bar\zeta)-
    Z(x_0^a,\zeta,\bar\zeta)
\end{equation}

\noindent Then, for every fixed value of $(\zeta,\bar\zeta)$, the surface
defined by the points $x^a$ that satisfy

\begin{equation}\label{eq:2}
    G(x_0^a,x^a,\zeta,\bar\zeta) =0
\end{equation}

\noindent is null and contains the point $x_0^a$, at which the vector
$\ell^a\equiv g^{ab}G,_b$ sweeps out the sphere of null directions as
$(\zeta,\bar\zeta)$ is allowed to vary. The parameters $(\zeta,\bar\zeta)$ are
thus coordinates on the observer's celestial sphere, and can
be taken as complex stereographic coordinates. Solving (\ref{eq:2}) for
$x^0\equiv t$ we have

\begin{equation}\label{eq:3}
    t = T(x_0^a,x^i,\zeta,\bar\zeta),
\end{equation}

\noindent namely, the value of the coordinate time of the point of intersection
of the family of null surfaces $G=0$ with the worldline $(x^i,t)$ of a static
source at the spatial position $x^i$, $i=1,2,3$. (This can be reformulated to
include a moving source.) Varying $(\zeta,\bar\zeta)$, different null surfaces
in the family $G=0$ are considered, all of which contain the observer $x_0^a$
and intersect the worldline of a potential source at $x^i$ at time
$T(x_0^a,x^i,\zeta,\bar\zeta)$. This time cannot be interpreted in terms of
curves joining the points $x_0^a$ and $(x^i,T)$ because there is an infinity of
curves lying on the surface $G=0$ joining the two points for which the
coordinate time of the end point is $t$: there is no way to pick a preferred
curve on the null surface.  In fact, among all possible curves joining the two
points on a given null surface $G=0$, there may even not be one that is {\it
geodesic}, and there are certainly many that become spacelike at least one
point.  Thus, even though $T(x_0^a,x^i,\zeta,\bar\zeta)$ does represent a time
function in a sense, we have no reason to identify it with the time of arrival
of a light signal.  However, $T(x_0^a,x^i,\zeta,\bar\zeta)$ is the time at
which a wavefront of the $(\zeta,\bar\zeta)-$null surface passes by the source
at $x^i$, so we can think that instead of looking at travel times along null
paths we are in effect looking at travel times of wavefronts from the observer
to the source. Both points of view are equivalent in the end, because
ultimately the values of $(\zeta,\bar\zeta)$ for which the intersection time
$T(x_0^a,x^i,\zeta,\bar\zeta)$ is stationary are null surfaces for which there
exists a null geodesic that joins $x_0^a$ with
$(x^i,T)$~\cite{arrival,fermatkling}.

Our recipe is: to extremize the function $T(x_0^a,x^i,\zeta,\bar\zeta)$
with respect to variations of $(\zeta,\bar\zeta)$ keeping $x^i$ fixed, and
interpret the extremum $T_e=T(x_0^a,x^i,\zeta_e,\bar\zeta_e)$ as the time
of arrival of a light signal from $x^i_0$ to $x^i$, giving an image in
the direction $(\zeta_e,\bar\zeta_e)$. These choices of $(\zeta,\bar\zeta)$
lead to null geodesics connecting the observer with the source. It follows that
the function $T(x_0^a,x^i,\zeta,\bar\zeta)$ can be viewed as a generalized
Fermat potential. On the other hand, since $T(x_0^a,x^i,\zeta,\bar\zeta)$,
in general will be given implicitly via (\ref{eq:2}), then extremizing
$T(x_0^a,x^i,\zeta,\bar\zeta)$ is equivalent to imposing the conditions

\begin{equation}\label{eq:4}
    G,_\zeta|_{x^a} = G,_{\bar\zeta}|_{x^a} = 0
\end{equation}

\noindent which are, in fact, the conditions for the envelope of the level
surfaces of $G$. This results from implicit differentiation
$\partial/\partial\zeta|_{x^i}$ of (\ref{eq:2}) where $t$
is assumed to be given by (\ref{eq:3}).  One obtains

\begin{equation}
    G,_\zeta|_{x^a} + G,_t T,_\zeta|_{x^i} = 0.
\end{equation}

\noindent Since $G,_t$ must be non-zero (otherwise it could not define $T$
implicitly), then $T,_\zeta|_{x^i}=0 \Leftrightarrow G,_\zeta|_{x^a}=0$.
Likewise with $T,_{\bar\zeta}|_{x^i}$. We refer to
$G(x_0^a,x^a,\zeta,\bar\zeta)$ as an ``implicit'' generalized Fermat potential,
making a slight abuse of terminology.

In fact, $G=0$ plus the envelope condition for $G$ as a joint set of equations
produces for us two things. 1.) A lens equation: an assignment of a value of
$(\zeta,\bar\zeta)$ on the celestial sphere for each source location $x^i$,
through light signals. 2.) A time of arrival of the light signal.  Thus, the
envelope conditions (\ref{eq:4}) on a level surface of $G$ embody our
generalized Fermat's principle: a version in terms of surfaces, rather than
paths.

Up to this point, we have assumed that the Fermat potential $G$ is somehow
known, and satisfies the eikonal equation.  Thus we have only considered the
kinematics of such implicit Fermat potentials: we have not considered how a
Fermat potential is affected by the presence of matter in the spacetime (the
dynamics). The matter (the deflector) is, of course, hidden away in the metric
$g_{ab}$ that determines the eikonal equation, so the eikonal constitutes part
of the dynamics, but the field equations for the metric itself are the Einstein
equations $R_{ab}-\frac12 g_{ab} R = T_{ab}$.  In principle, one could first
solve for the metric, and then use the metric in the eikonal to obtain the
Fermat potential $G$. There is no general solution to the Einstein equations in
terms of data or matter sources, because of their nonlinearity, therefore this
route does not have the potential to provide us with a one-step dynamical
equation for the Fermat potential driven directly by the matter source, which
would be analogous to Eq.~(\ref{eq:0}).

Notwithstanding, there is an approach to general relativity which has the
potential to directly lead to dynamical equations for implicit Fermat
potentials~\cite{nsf1,nsflin,nsflor,nsfdyn}. We describe the resulting dynamics
of implicit Fermat potentials through such an approach in Section~\ref{sec:2}.
An application of the dynamics to the case of weak fields is worked out in
Section~\ref{sec:3}.  The main results of the application to weak fields are
two lens equations: one for the case of thick compact objects and the other one
for the case of lensing by gravitational waves.  Whether these equations can be
of use in astrophysical situations remains to be seen, but our point is that
they can be written down in principle.

\section{Dynamics of the implicit Fermat potential}\label{sec:2}

The formulation of general relativity via null surfaces~\cite{nsf1,nsfdyn}
provides a setting for discussing the dynamics of implicit Fermat
potentials in the sense of the previous Section. This formulation is
described in complete detail in~\cite{nsf1,nsfdyn}. Here we provide a
(necessarily brief) summary of the main aspects of the formulation, hoping
that the unfamiliar reader is able to complement this Section
with~\cite{nsf1,nsfdyn}.

In the null-surface approach, the aim is to rewrite the Einstein
equations in terms of variables that are to be thought of as more
fundamental than the metric $g_{ab}$, which can then be generated
from them.  The fundamental objects of the formulation are two
functions of six variables, $Z(x^a,\zeta,\bar\zeta)$ and
$\Omega(x^a,\zeta,\bar\zeta)$.  The variables $x^a$ represent
points in a four-dimensional spacetime with a Lorentzian metric to
be determined. The variables $(\zeta,\bar\zeta)$ represent a
sphere fiber at every point $x^a$; eventually $(\zeta,\bar\zeta)$
acquires the meaning of the sphere of null directions at each
spacetime point. The first function, $Z$, whose level surfaces are
null for all values of $(\zeta,\bar\zeta)$ encodes the conformal
structure of the spacetime. That is to say that if the metric of
the spacetime is known, then $Z(x^a,\zeta,\bar\zeta)$ satisfies
the eikonal equation $g^{ab}Z,_aZ,_b=0$ for all values of
$(\zeta,\bar\zeta)$.  The second function, $\Omega$, represents
the scale (or conformal) factor of the spacetime with conformal
structure given by $Z$ (all conformally related spacetimes have
the same conformal structure; the scale factor breaks the
conformal invariance). From the two variables, $Z$ and $\Omega$, a
Lorentzian metric can easily be constructed. Field equations,
equivalent to the Einstein equations can be imposed on them.

These field equations for $Z$ and $\Omega$ are in general quite
complicated. However, for the weak field applications that we will
treat they become quite simple and are easily reduced to
quadratures. We  will first present the full equations before
giving the linearization.

It is perhaps easiest if the equations are expressed in two steps. We first
introduce several auxiliary variables obtained from derivatives of
$Z(x^a,\zeta,\bar\zeta)$; we define four variables
$\theta^i(x^a,\zeta,\bar\zeta)$ which can be considered as coordinates given
intrinsically by the families of null surfaces by

\begin{mathletters}\label{theta}
\begin{eqnarray}
    \theta^0 &\equiv & Z \equiv u,  \\
    \theta^+ &\equiv & \eth Z \equiv \omega, \\
     \theta^- &\equiv& \overline\eth Z \equiv \overline\omega,\\
    \theta^1 &\equiv& \eth\overline\eth Z \equiv R.
\end{eqnarray}
\end{mathletters}

\noindent Using these relations we can go back and forth between
$x^a$ and $\theta^i$. Further, it is assumed that these relations
can be inverted locally so that we have

\begin{equation}
    x^a=x^a(\theta^i,\zeta,\bar\zeta).  \label{inverse}
\end{equation}

\noindent In addition to $\theta^i$, we define another fundamental
variable (and its complex conjugate) by

\begin{equation}
    \Lambda(\theta^i,\zeta,\bar\zeta)=\eth^2Z,  \label{lambda}
\end{equation}

\noindent where the $x^a$ on the right-hand side have been replaced by the
$\theta^i$ via Eq.~(\ref{inverse}).

For the second step, the Einstein equations are written in terms of
$\Lambda(\theta^i,\zeta,\bar\zeta)$ and $\Omega(\theta^i,\zeta,\bar\zeta)$ as

\begin{mathletters}\label{nsfequations}
\begin{eqnarray}
    \Omega,_{11} &=& Q(\Lambda) \Omega + T\Omega^3,    \label{einst1}\\
    \widehat{\eth}\Omega &=& \frac12 W(\Lambda) \Omega ,    \label{einst2} \\
    M(\Lambda) &=& 0,                   \label{einst3}
\end{eqnarray}
\end{mathletters}

\noindent where we have used the notation
${},_i=\frac{\partial}{\partial\theta^i}$ and in particular

\begin{equation}
    \Omega,_{11}=\frac{\partial^2\Omega}{\partial R^2}.  \label{R}
\end{equation}

\noindent The quantity $T$ encodes the information in the stress-energy tensor
$T_{ab}$, while $(Q,W,M)$ are explicit functions of $\Lambda $ and its
derivatives:

\begin{mathletters}\label{nsfsymbols}
\begin{eqnarray}
    Q &\equiv & \frac{1}{4q} \left(\Lambda,_{11}\overline\Lambda,_{11}
         +\frac{3}{2q} (q,_1)^2
         -q,_{11}\right),    \label{Q}\\
    M &\equiv& \widehat{\eth} (\Lambda,_1) -2\Lambda,_-
            -\Big( W +\widehat{\eth} \ln q
            \Big) \Lambda,_1  ,  \label{M} \\
    W &\equiv&   \left( \Lambda,_+
                + \frac12\overline{\widehat{\eth}}(\Lambda,_1)
                + \frac12 \Lambda,_1\overline\Lambda,_-
                + \frac14 \Lambda,_1\widehat{\eth}(\overline\Lambda,_1)
                - \frac12  \widehat{\eth}\ln q
                - \frac14  \Lambda,_1\overline{\widehat{\eth}}\ln q\right)
                    \left(1-\frac14\Lambda,_1\overline\Lambda,_1\right)^{-1},
    \label{W}
\end{eqnarray}
\end{mathletters}

\noindent with

\begin{equation}\label{q}
q \equiv 1-\Lambda,_1\overline\Lambda,_1.
\end{equation}

\noindent The operator $\widehat{\eth}$ is the usual $\eth$ with $x^a$ held
constant; but operating on functions $f(\theta^i,\zeta,\bar\zeta)$ it takes the
form

\begin{equation}\label{edth-hat}
\widehat{\eth} f(\theta^i,\zeta,\bar\zeta)
    = \eth f|_{\theta^i}
     + \theta^+ f,_0
     + \Lambda f,_+
     + \theta^1 f,_-
     + \Big( \frac{\Lambda,_1\overline J + J}
                  {1-\Lambda,_1\overline\Lambda,_1}
     \Big)f,_1
\end{equation}

\noindent with

\begin{equation}
       J =  - 2     \theta^+
        +   \theta^-   \Lambda,_0
        + \overline\eth \Lambda|_{\theta^i}
        + \theta^1          \Lambda,_+
        +\overline      \Lambda\Lambda,_-
\end{equation}

\noindent For the purposes of applying the  $\widehat{\eth}$ operator,
$\Lambda$ and  $\Omega$ have spin weights 2 and 0.

Eq.~(\ref{einst1}) is a direct translation of the Einstein equation,
$G_{ab}=T_{ab}$, while Eqs.~(\ref{einst2}) and (\ref{einst3}), referred to as
the metricity conditions, are the requirements that a unique metric can be
constructed from $Z$ and $\Omega$.

The field equations couple $\Lambda $ and $\Omega $ in a complicated non-linear
manner that makes finding exact solutions an almost impossible task.
Nevertheless, as we will shortly see, the linearization becomes quite simple.
In principle, the procedure for finding the Fermat potential would be to solve
Eqs.~(\ref{nsfequations}) for $\Lambda =\Lambda(\theta^i,\zeta,\bar\zeta)
=\Lambda(Z,\eth Z,\overline\eth Z,\eth\overline\eth Z,\zeta,\bar\zeta)$ and then return
to Eq.~(\ref{lambda}), written in the form

\begin{equation}\label{lambda2}
    \eth^2 Z
    = \Lambda(Z, \eth Z, \overline\eth Z, \eth\overline\eth Z,\zeta,\bar\zeta)
\end{equation}

\noindent so that it becomes a dynamical equation for $Z(x^a,\zeta,\bar\zeta)$
with a known right-hand side. Actually, since $\Lambda $ is a complex function,
(\ref{lambda2}) is a pair of over-determined second-order PDEs for $Z$ in the
two variables $(\zeta,\bar\zeta)$. From general considerations~\cite{niky}, the
solution space is four-dimensional; the solutions have four constants of
integration, $x^a$, that become the spacetime coordinates. Eq.~(\ref{lambda2})
is our dynamical equation for the Fermat potential.  In this regard, we point
out that Eq.~(\ref{einst3}) is precisely the vanishing of the generalized
Wunschmann invariant that is central to the problem of the classification of
this type of overdetermined systems of PDE's~\cite{niky,diffgeomdiffeqs}.

Some effort has been devoted to the issue of rewriting the field equations in
terms of $Z$ and $\Omega$ by eliminating $\Lambda$ via (\ref{lambda2}) and
(\ref{theta}), with partial success~\cite{scri}. To do this exactly appears to
be a formidable task. Nevertheless, in the case of weak fields, the
linearization of the field equations allows us to combine the content of the
field equations (\ref{nsfequations}) and of (\ref{lambda2}) into a single
dynamical equation for $Z$ in terms of given data~\cite{nsfdyn}. We describe,
in the following section, this application of our program to weak fields.

\section{Lens equations in the case of weak fields} \label{sec:3}

The field equations (\ref{nsfequations}) have input of two types. In the first
place, there is the matter and stresses, encoded into the symbol $T$.
Additionally, there is also gravitational radiation at the boundaries of the
spacetime that must be considered free input. This is because the field
equations are equivalent to the characteristic problem for the Einstein
equations, in which case the gravitational radiation incoming into the
spacetime must be specified~\cite{nsflin,scri}, usually in the form of the
complex shear $\sigma(u,\zeta,\bar\zeta)$of a null congruence of the Bondi
type evaluated at the boundary
of the spacetime, ${\cal J}^-$. This is a complex function of
three arguments, the local Bondi coordinates on ${\cal J}^-$.

We concentrate on the case in which all input sources or data are
small, namely, there exists a smallness parameter $\epsilon$ by
which the magnitude of the source and data can be measured, and
such that its powers can be neglected. This will lead to a
linearized version of the dynamics of the Fermat potentials
discussed in the previous section. However, within this regime
there are two distinct main cases of interest.  One is the case
where the spacetime is empty of matter and the gravitational
radiation is small, that is: $T=0$ and $|\sigma|\sim \epsilon$.
We refer to this case as the {\it gravitational-radiation case}.  The
other case is such that the matter source is weak but finite,
and the gravitational radiation is negligible, namely: $|T| \sim
\epsilon$ and $|\sigma|\sim \epsilon^k$ with $k\ge 2$. This case
will be referred to as the {\it near-Newtonian case}.

\subsection{The gravitational-radiation case}

We assume that we have a purely radiative spacetime, in which the
stress-energy tensor vanishes and the gravitational radiation is
small ($T=0$ and $|\sigma|\sim\epsilon$). The spacetime is then
flat up to a small perturbation.  In flat spacetime, null planes
in all possible directions constitute a 2-parameter family of
solutions to the eikonal equation, leading to $Z_{\mbox{\small
flat}}(x^a,\zeta,\bar\zeta) =x^a\ell_a$ where

\begin{equation}\label{ell}
\ell^a \equiv
\frac{1}{\sqrt{2}(1+\zeta \bar\zeta)}(1+\zeta \bar\zeta,\zeta +\bar\zeta,
-i(\zeta -\bar\zeta),\zeta\bar\zeta-1)
\end{equation}

\noindent is a future directed null vector
pointing in the $(\zeta,\bar\zeta)$-direction.  We may expect to have then

\begin{equation}
    Z(x^a,\zeta,\bar\zeta)
    = x^a\ell_a + \widetilde{Z}(x^a,\zeta,\bar\zeta)
\end{equation}

\noindent with $|\widetilde{Z}|\sim \epsilon$. Since $\eth^2\ell^a = 0$, then
the variable $\Lambda$ in the field equations (\ref{nsfequations}) is of
first order in $\epsilon$, and products of $\Lambda$ and its derivatives
can be neglected.  Therefore the field equations (\ref{nsfequations}) can
be linearized with respect to $\Lambda$ around $\Lambda=0$. They become

\begin{mathletters}\label{nsfequationslin}
\begin{eqnarray}
    \Omega,_{11} &=& 0,    \label{nsfeinsteinlin}\\
    \widehat{\eth}\Omega
    &=& \frac12 \left(\Lambda,_+
                 +\frac12\overline{\widehat{\eth}}(\Lambda,_1)\right) \Omega ,
                   \label{nsfmesslin} \\
    2\Lambda,_- -\widehat{\eth} (\Lambda,_1)&=& 0.
    \label{nsfmetricitylin}
\end{eqnarray}
\end{mathletters}

\noindent By Eq.~(\ref{nsfeinsteinlin}) with boundary conditions consistent
with an asymptotically flat spacetime (in which $\Omega\to 1$ along null
directions), we must have $\Omega=1$.  Thus (\ref{nsfmesslin}) becomes

\begin{equation} \label{nsfmesslinvac}
    2\Lambda,_+
         +\overline{\widehat{\eth}}(\Lambda,_1)
    =0
\end{equation}

\noindent Equations (\ref{nsfmetricitylin}) and (\ref{nsfmesslinvac}) can
be manipulated through algebra, differentiation and integration (with
boundary conditions consistent with an asymptotically flat spacetime), to
yield~\cite{nsflin}

\begin{equation}\label{thiseqn}
    \overline{\widehat{\eth}}^2\Lambda =
    \overline{\widehat{\eth}}{}^2\sigma(\theta^0,\zeta,\bar\zeta)
    +    \widehat{\eth}^2\overline\sigma(\theta^0,\zeta,\bar\zeta).
\end{equation}

\noindent Using (\ref{theta}) and (\ref{lambda2}), this equation can be
rewritten as an equation for $Z$:

\begin{equation}\label{lccvac}
    \overline{\eth}^2\eth^2 Z =
    \overline{\widehat{\eth}}{}^2\sigma(x^a\ell_a,\zeta,\bar\zeta)
    +    \widehat{\eth}^2\overline\sigma(x^a\ell_a,\zeta,\bar\zeta).
\end{equation}

\noindent Notice that in (\ref{lccvac}) the operator $\eth$ acts
at fixed $x^a$, so that all three arguments in
$\sigma(x^a\ell_a,\zeta,\bar\zeta)$ contribute to the
eth-derivative.  In order to obtain this equation from
(\ref{thiseqn}), $\sigma(\theta^0,\zeta,\bar\zeta)$ is
approximated by $\sigma(x^a\ell_a,\zeta,\bar\zeta)$, an
approximation that is justified because $\sigma$ is a first-order
quantity, therefore first-order terms in its arguments can be
neglected. Since $Z$ leads to the Fermat potential $G$ in a
trivial manner, we consider this the dynamical equation for the
implicit Fermat potential in the presence of weak gravitational
radiation. The operator $\overline{\eth}^2\eth^2$ is the double
Laplacian on the sphere $(\zeta,\bar\zeta)$, whereas the
right-hand side acts as a known source. This is the analog of
Eq.~(\ref{eq:0}) for lensing by gravitational waves without the
approximations of thin lenses or small angles.

There exists a simple Green function~\cite{ivancovich} for the double
Laplacian $\overline{\eth}^2\eth^2$, given by

\begin{equation}\label{green}
    F(\zeta,\bar\zeta,\zeta',\bar\zeta')
    =
    \frac{1}{4\pi}
    \ell'_a\ell^a \ln (\ell'_a\ell^a),
\end{equation}

\noindent which can be used to express the general solution to
(\ref{lccvac}) in closed form in terms of the source:

\begin{equation}\label{Zvlosedvac}
    Z(x^a,\zeta,\bar\zeta)
    = x^a\ell_a
    +\frac{1}{4\pi}\int_{S^2}
    \Big(\overline{\widehat{\eth}}{}'{}^2\sigma(x^a\ell'_a,\zeta',\bar\zeta')
    +    \widehat{\eth}'{}^2\overline\sigma(x^a\ell'_a,\zeta',\bar\zeta')
    \Big)\ell'_a\ell^a \ln (\ell'_a\ell^a)
    dS^2{}'.
\end{equation}

\noindent This expression for $Z$ can now be used to construct the implicit
Fermat potential $G=Z(x^a,\zeta,\bar\zeta)-Z(x_0^a,\zeta,\bar\zeta)$.  The
two envelope conditions $\eth G = \overline\eth G = 0$  then take the
form

\begin{eqnarray}\label{lensgravwaves}
    0
    &=& (x^a-x_0^a)\eth\ell_a   \nonumber\\
    &&+\frac{1}{4\pi}\int_{S^2}
    \Big(\overline{\hat{\eth}}{}'{}^2
        [(\sigma(x^a\ell'_a,\zeta',\bar\zeta')
         -\sigma(x_0^a\ell'_a,\zeta',\bar\zeta')]\nonumber\\
    &&
       \hspace{1cm} +\widehat{\eth}'{}^2
        [\overline\sigma(x^a\ell'_a,\zeta',\bar\zeta')
        -\overline\sigma(x_0^a\ell'_a,\zeta',\bar\zeta')]
    \Big)\eth[\ell'_a\ell^a \ln (\ell'_a\ell^a)]
    dS^2{}'.
\end{eqnarray}

\noindent By our discussion in Section~\ref{sec:1}, Equations
(\ref{lensgravwaves}) and
$Z(x^a,\zeta,\bar\zeta)-Z(x_0^a,\zeta,\bar\zeta)=0$ combined are
equivalent to the lens mapping and time delay for the case of
lensing by weak gravitational waves without the approximations of
thin lenses or small angles. Since there is no lens plane in this
case, the lens equation is interpreted as a mapping from the
observer's celestial sphere to the source's spatial location.  The
lens mapping is understood in the following sense. We assume a
spin-2 function $\sigma(u,\zeta,\bar\zeta)$ is given, perhaps
constructed with some model in mind. This function represents the
gravitational radiation that reaches the boundary of an
asymptotically flat vacuum spacetime.  We can specify our
coordinates by setting the origin on the observer's worldline, so
$x_0^a=(\tau,0,0,0)$. The point $x^a=(t,x,y,z)$ that satisfies
both equations $Z(x^a,\zeta,\bar\zeta)-Z(x_0^a,\zeta,\bar\zeta)=0$
and (\ref{lensgravwaves}) then lies on the observer's lightcone,
and is reached by a null geodesic in the direction
$(\zeta,\bar\zeta)$ at the observer. These are three equations for
the four variables $x^a$, so one of them will be free.  As usual,
we assume that this free coordinate represents a distance to the
source, or can be obtained from a measured distance to the source.
The remaining two spatial coordinates then specify the angular
location of the source, and the time coordinate represents the
travel time of the light signal from the source to the observer.

It may not be a simple matter to solve for the two angular
coordinates and time from
$Z(x^a,\zeta,\bar\zeta)-Z(x_0^a,\zeta,\bar\zeta)=0$ and
(\ref{lensgravwaves}), in order to express the lens mapping in
closed form even if a function $\sigma$ is given explicitly. In
fact, it is not even clear that the lens mapping can be expressed
in closed form. But the point that we wish to make is that in
principle, the lens mapping is there. Work on an explicit example
of this lens mapping is in progress and will be reported
elsewhere.

What would be the use of lensing by gravitational radiation?
Lensing by extremely compact objects provides a means to study the
mass structure of the deflector. Likewise, we can think of using
lensing by gravitational waves to infer information about the
gravitational waves themselves. In principle, lensing events are
natural detectors of gravitational waves. In practice, however,
these lensing events are expected to lie below the observational
limit, and be overshadowed by the magnitude of lensing effects by
compact objects.


\subsection{The near-Newtonian case}

We turn our attention now to the opposite case, namely: there are sources
of weak strength but there is little or no incoming gravitational
radiation. This would include, for instance, the case of static spacetimes
with weak (isolated) matter sources, or any matter sources that are static
in first approximation. We thus have, by assumption, $|T|\sim \epsilon$
and $|\sigma|\sim \epsilon^k$ with $k\ge 2$. The source $T$ is a function of
six variables obtained from the stress energy tensor $T_{ab}$ by

\begin{equation}
    T(\theta^i,\zeta,\bar\zeta) \equiv T_{ab}(x^a)\ell^a\ell^b,
\end{equation}

\noindent where
\begin{equation}
    x^a = (2\theta^0 +\theta^1)\ell^a
        +\theta^0\eth\overline\eth\ell^a
        -\theta^-\eth\ell^a
        -\theta^+\overline\eth\ell^a ,
\end{equation}

\noindent where $\ell^a$ is given by (\ref{ell}). In this case,
the Einstein equation (\ref{einst1}) splits into two hierarchical
equations, one for the zeroth-order term $\Omega_0$ and the other
one for the first-order term $\tilde{\Omega}$:

\begin{mathletters}
\begin{eqnarray}
    (\Omega_0),_{11} &=& 0,     \\
    \widetilde{\Omega},_{11} &=& T\Omega_0.
\end{eqnarray}
\end{mathletters}

\noindent It has been shown~\cite{nsflin} that in the asymptotically flat
regime, one can take $\Omega_0 = 1$ consistently with Eq.~(\ref{einst2})
up to first order. Thus we take $\Omega_0 = 1$ and the field equations
(\ref{nsfequations}) become

\begin{mathletters}\label{nsfequationsnewt}
\begin{eqnarray}
    \Omega,_{11} &=& T,    \label{nsfeinsteinewt}\\
    \widehat{\eth}\Omega
    &=& \frac12 \left(\Lambda,_+
    +\frac12\overline{\widehat{\eth}}(\Lambda,_1)\right)  ,\label{nsfmessnewt} \\
    2\Lambda,_- -\widehat{\eth} (\Lambda,_1)&=& 0.
    \label{nsfmetricitynewt}
\end{eqnarray}
\end{mathletters}

\noindent where we have used the fact that also in this case the
spacetime can be represented as a small perturbation off flat
space, in which case $\Lambda$ is of first-order in $\epsilon$. By
virtue of Eq.~(\ref{nsfeinsteinewt}), the conformal factor
$\Omega$ can be obtained from the source by quadratures:

\begin{equation}\label{omeganewt}
    \Omega = 1 - \int_{\theta^1}^\infty
             \int_{\theta^1{}'}^\infty T d\theta^1{}'d\theta^1{}''
\end{equation}

\noindent where the upper limit of the integral has been chosen at
$\theta^1=\infty$, the null boundary of the asymptotically flat spacetime.
Thus $\Omega$ can be considered as a given source of
Eq.~(\ref{nsfmessnewt}).  Equations (\ref{nsfmessnewt}) and
(\ref{nsfmetricitynewt}) can be manipulated through algebra,
differentiation and integration~\cite{nsflin}, to yield

\begin{equation}\label{lcclambdanewt}
    \overline{\widehat{\eth}}^2\Lambda
    =
    \int^\infty_{\theta^0}\left(
 3\overline\eth^2\int^\infty_{\theta^1} (\widehat{\eth}\Omega),_- d\theta^1{}'
+3\eth^2\int^\infty_{\theta^1} (\overline{\widehat{\eth}}\Omega),_+ d\theta^1{}'
    -2\widehat{\eth}\overline{\widehat{\eth}}(2\Omega
    -\widehat{\eth}\overline{\widehat{\eth}}\Omega)
            \right)
    d\theta^0{}'.
\end{equation}

\noindent The upper limit $\theta^0=\infty$ represents the point $i^0$ at the
null boundary of an asymptotically flat spacetime (timelike infinity).

Using (\ref{theta}) and (\ref{lambda2}),
Eq.~(\ref{lcclambdanewt}) can be rewritten as an equation for $Z$:

\begin{equation}\label{lccnewt}
    \overline{\eth}^2\eth^2 Z
    =
    \int_{\theta^0_0}^{x^a\ell_a}\left(
3\overline\eth^2\int_\infty^{x^a\eth\overline\eth\ell_a}
                     (\widehat{\eth}\Omega),_- d\theta^1{}'
+3\eth^2\int_\infty^{x^a\eth\overline\eth\ell_a}
                     (\overline{\widehat{\eth}}\Omega),_+ d\theta^1{}'
    -2\widehat{\eth}\overline{\widehat{\eth}}(2\Omega
    -\widehat{\eth}\overline{\widehat{\eth}}\Omega)
            \right)
    d\theta^0{}'.
\end{equation}

\noindent This is a Poisson-type equation, since the operator in
the left-hand side is the double Laplacian on the sphere
$(\zeta,\bar\zeta)$, and the right-hand side is a given source, by
virtue of (\ref{omeganewt}). Again, since $Z$ leads to the Fermat
potential $G$ in a trivial manner, we consider this the dynamical
equation for the Fermat potential. This equation thus  plays a
role analogous to (\ref{eq:0}) and generalizes Eq.~(\ref{eq:0}) by
removing the approximations of thin lenses and small angles.

We can now proceed to write down the lens mapping explicitly.  We
start by writing down the solution of (\ref{lccnewt}) in terms of the
source in the right-hand side, by means of the Green function
$F(\zeta,\bar\zeta,\zeta',\bar\zeta')$ given by Eq.~(\ref{green}):

\begin{equation}
    Z(x^a,\zeta,\bar\zeta)
    = x^a\ell_a
    +\frac{1}{4\pi}\int_{S^2}
    S(x^a,\zeta',\bar\zeta')\ell'_a\ell^a \ln (\ell'_a\ell^a)
    dS^2{}',
\end{equation}

\noindent with

\begin{equation}\label{S}
    S(x^a,\zeta,\bar\zeta)
    \equiv
    \int^{\infty}_{x^a\ell_a}\left(
3\overline\eth^2\int^\infty_{x^a\eth\overline\eth\ell_a}
     (\widehat{\eth}\Omega),_- d\theta^1{}'
+3\eth^2\int^\infty_{x^a\eth\overline\eth\ell_a}
     (\overline{\widehat{\eth}}\Omega),_+ d\theta^1{}'
-2\widehat{\eth}\overline{\widehat{\eth}}(2\Omega
-\widehat{\eth}\overline{\widehat{\eth}}\Omega)
            \right)
    d\theta^0{}'.
\end{equation}

\noindent We can now  construct the Fermat potential
$G=Z(x^a,\zeta,\bar\zeta)-Z(x_0^a,\zeta,\bar\zeta)$ and write down the two
envelope conditions $\eth G = \overline\eth G = 0$, which then take
the form

\begin{equation}\label{lensnewt}
    0
    = (x^a-x_0^a)\eth\ell_a
    +\frac{1}{4\pi}\int_{S^2}
    S(x^a,\zeta',\bar\zeta')
    \eth[\ell'_a\ell^a \ln (\ell'_a\ell^a)]
    dS^2{}'.
\end{equation}

\noindent This equation, with
$Z(x^a,\zeta,\bar\zeta)-Z(x_0^a,\zeta,\bar\zeta)=0$, is a generalization of the
astrophysical lens equation and time delay that removes the thin-lens and
small-angle approximations. As in the previous subsection, there is no lens
plane, and the mapping takes points on the observer's celestial sphere into the
source's spatial locations. To be more specific, we can choose the coordinates
so that the observer lies at the origin $x_0^a=(\tau,0,0,0)$. The stress-energy
tensor $T_{ab}$ of a given matter configuration on a flat background is assumed
to be given, perhaps with some model in mind. The quantity
$T=T_{ab}\ell^a\ell^b$ is constructed, and is used as the source of the
quadrature for the quantity $\Omega$ via Eq.~(\ref{omeganewt}). The quantity
$\Omega$ thus obtained is used to obtain $S$ according to (\ref{S}). The
combined equations $Z(x^a,\zeta,\bar\zeta)-Z(x_0^a,\zeta,\bar\zeta)=0$ and
(\ref{lensnewt}) provide three equations for the four variables $x^a$.  The
values of $x^a$ that satisfy the three equations lie on the observer's
lightcone. One of the four variables $x^a$ remains free, and is assumed to be
given in terms of some observed physical distance.  The remaining three
variables represent the angular location of the source and the time of arrival
of the light signal.

It may be complicated to write down the lens mapping in closed
form, even if $T_{ab}$ is given explicitly.  However, one can
envision a physical situation that is close to a thin lens, in
which case we would expect this scheme to return a kind of
``post-thin'' lens equation. In other words, it is worth
considering the possibility of obtaining corrections to the
thin-lens approach by means of this scheme. The effect of the
thickness of the lens has only rarely been considered in the
literature~\cite{hammer,padmanabhan,thick,pyne1,pyne2}.

\section{Remarks and outlook}

By setting up a framework within the null--surface formulation to discuss
the dynamics of Fermat potentials for gravitational lensing in full
generality, we have achieved two main goals, as argued in the following.

In the first place, we have provided two lens equations that
generalize the astrophysical approach to lensing in two ways: 1.)
both lens equations remove the approximations of thin lenses and
small angles; and 2.) the lens equations include the case of
lensing by gravitational waves.  Even though the exceeding
complexity of the case of non-vanishing matter sources --
Eq.~(\ref{lensnewt}) -- puts its applicability into question, the
point of our work is to show that, in principle, the thin-lens
approximation can be removed in a meaningful manner.  On the other
hand, the obvious simplicity of the case of vanishing matter
sources -- Eq.~(\ref{lensgravwaves}) -- opens a door to the
phenomenon of lensing by gravitational waves, which has
occasionally been treated before, within the framework of the
thin-lens approximation~\cite{faraoniwaves} and in a statistical
framework~\cite{jaffekaiser}. In principle, both cases could be
combined in a straightforward manner, simply by adding their
respective $Z$ functions, therefore the spectrum of many possible
applications to weak fields is opened in this work.

One might question the necessity of the use of any type of
Fermat's principle when one moves away from the thin-lens
approximation on the basis that the direct integration of the null
geodesics on a linearized perturbation off flat space would
clearly yield a lens map. In \cite{pyne1}, for instance, the
equations for the null geodesics of a linearized perturbation off
flat space are written down explicitly and integrated to obtain
the bending angle of a moving thin lens, whereas in~\cite{pyne2}
the direct integration of the null geodesics of a perturbation off
an isotropic cosmology is used to derive the cosmological
thin-lens equation without the use of broken paths. Still, so far
as we are aware, very rarely if ever has the direct integration of
the null geodesics led to a scheme in any way comparable with the
thin-lens scheme in practicality. Ideally, one would like to have
a correction to the thin-lens scheme that represents the effects
of the width of the lens and an estimate of the regime in which
such corrections cannot be neglected. One can easily argue that
the approach that we present here may have the potential to make a
contribution towards such a goal and deserves to be pursued. The
usefulness of our approach must first be tested in particular
cases where results are already available. In this respect, the
thin-lens scheme has been easily reproduced in our
approach~\cite{fermatkling}, including the case of moving lens
planes. It would be of great interest to be able to relate our
lens equations to existing results involving lensing by
gravitational waves, such as~\cite{jaffekaiser} among others. The
reader should be aware that, because of the statistical nature of
the problem considered in \cite{jaffekaiser}, our equation
(\ref{lensgravwaves}) cannot be compared directly with results in
that article.

One should keep in mind, as well, that the treatment in this work
is limited in at least two senses. We have only considered weak
fields as perturbations off flat space. Therefore, the application
to isotropic cosmologies is not covered here. There are reasons to
think that the scheme developed here can be adapted to isotropic
cosmologies with little effort. Work on this line is in progress
and will be reported elsewhere. Secondly, we have only considered
a coordinate-based treatment of lensing, as opposed to an
optical-distance treatment. By virtue of this we have stayed away
from a great technical and practical difficulty in observational
lensing, which is the use of a meaningful observable distance to
the source. We have bypassed this difficulty by simply using
abstract coordinates.  The use of an observable coordinate
distance to the source would greatly complicate the lens equations
obtained here in form, but it would not affect their underlying
content.

In second place, we have here demonstrated that the null-surface
formulation of general relativity has the potential for physical
applications of interest, in spite of its apparent level of
abstraction.

\acknowledgments

This work was supported by the NSF under grant No. PHY-0070624 to
Duquesne University, and  No. PHY-0088951 to the University of Pittsburgh.


\begin{thebibliography}{10}

\bibitem{EFS}
P. Schneider, J. Ehlers, and E.~E. Falco, {\em Gravitational
Lenses} (Springer-Verlag, New York, 1992).

\bibitem{pettersbook}
A.~O. Petters, H. Levine, and J. Wambsganss, {\em Singularity
theory and gravitational lensing} (Birkh\"{a}user, Boston, 2001).

\bibitem{fermatkling}
S. Frittelli, T.~P. Kling, and E.~T. Newman, Fermat potentials in
  non-perturbative gravitational lensing, preprint.

\bibitem{arrival}
S. Frittelli and E.~T. Newman,  in {\em The Universe}, edited by
N. Dadhich and A. Kembhavi (Kluwer Academic Publishers, Dordrecht,
2000), \texttt{gr-qc/9810067}.

\bibitem{nsf1}
S. Frittelli, C.~N. Kozameh, and E.~T. Newman, J. Math. Phys. {\bf
36}, 4984 (1995).

\bibitem{nsflin}
S. Frittelli, C.~N. Kozameh, and E.~T. Newman, J. Math. Phys. {\bf
36}, 5005 (1995).

\bibitem{nsflor}
S. Frittelli, C.~N. Kozameh, and E.~T. Newman, J. Math. Phys. {\bf
36}, 4975 (1995).

\bibitem{nsfdyn}
S. Frittelli, C.~N. Kozameh, and E.~T. Newman, J. Math. Phys. {\bf
36}, 6397 (1995).

\bibitem{niky}
S. Frittelli, N. Kamran, and E.~T. Newman, Differential equations
and conformal geometry, accepted for publication in the Journal of
Geometry and Physics, October 2001.

\bibitem{diffgeomdiffeqs}
S. Frittelli, C. Kozameh, and E.~T. Newman, Commun. Math. Phys.
{\bf 223}, 383 (2001).

\bibitem{scri}
S. Frittelli, C.~N. Kozameh, and E.~T. Newman, Phys. Rev. D {\bf
56}, 4729 (1997).

\bibitem{ivancovich}
J. Ivancovich, C.~N. Kozameh, and E.~T. Newman, {J}. Math. Phys.
{\bf 30}, 45 (1989).

\bibitem{hammer}
F. Hammer, Astron. Astroph. {\bf 144}, 408 (1985).

\bibitem{padmanabhan}
T. Padmanabhan and K. Subramanian, Mon. Not. R. Astr. Soc. {\bf
233}, 265 (1988).

\bibitem{thick}
S. Frittelli and T.~E. Oberst, Phys. Rev. D {\bf 65}, 023005 (2001).

\bibitem{pyne1} T. Pyne and M. Birkinshaw, Ap. J. {\bf 415}, 459 (1993).

\bibitem{pyne2} T. Pyne and M. Birkinshaw, Ap. J. {\bf 458}, 46 (1996).

\bibitem{faraoniwaves}
V. Faraoni, Int. J. Mod. Phys. D {\bf 7}, 409 (1998).

\bibitem{jaffekaiser}
N. Kaiser and A. Jaffe,  Ap. J. {\bf 484}, 545 (1997).


\end{thebibliography}

\end{document}